\documentclass[a4paper,11pt]{article}
\pdfoutput=1 

\usepackage{jheparxiv} 

\usepackage[T1]{fontenc} 

\usepackage{xcolor}
\usepackage{tensor}
\usepackage{amsmath}
\usepackage{amssymb}
\usepackage{mathrsfs}
\usepackage{graphicx}
\usepackage{stmaryrd}
\usepackage{twistor}
\usepackage{commath}
\usepackage{xcolor}
\usepackage{mathtools}
\usepackage{hyperref}

\title{\Large{On Carrollian Loop Amplitudes for Gauge Theory and Gravity}}

\author[a]{Vijay Nenmeli}
\author[b]{\& Bin Zhu}

\affiliation[a]{School of Mathematics and Maxwell Institute for Mathematical Sciences \\
University of Edinburgh, EH9 3FD, U.K.}

\affiliation[b]{School of Physics, Nankai University, Weijin Road 94, Tianjin 300071, P.R. China}

\emailAdd{v.v.nenmeli@sms.ed.ac.uk}
\emailAdd{bzhu@nankai.edu.cn}

\begin{document} 

\abstract{Carrollian amplitudes are scattering amplitudes of massless particles written in position space at null infinity. We study various aspects of Carrollian amplitudes for gauge theory and gravity at loop level using primarily the modified Mellin prescription of~\cite{Bagchi:2022emh}. Finite one-loop four-point Carrollian amplitudes in gauge theory are shown to maintain an analytic structure similar to tree level  results. We compute the one-loop four-point  Carrollian MHV amplitudes in planar $\mathcal{N}=4$ super Yang-Mills theory, which are expressed as differential operators acting on tree level Carrollian amplitudes. This result is generalized to all loop orders using the Bern-Dixon-Smirnov (BDS) formula. Similar structures are observed at one-loop for Carrollian MHV amplitudes in $\mathcal{N}=8$ supergravity. We next consider $2\to 2$ scattering of massless scalars via gravitational interactions in the eikonal regime and show that the corresponding Carrollian amplitudes exhibit logarithmic behavior in the `Carroll time' $u$. We compute the discontinuities of these Carrollian amplitudes up to $\mathcal{O}(G^3)$ and show that they are descendants of Carrollian Born amplitudes. We observe similar logarithmic behavior in Carrollian amplitudes associated with the one-loop scalar box diagram. The dependence of this amplitude on dual scaling dimensions also differs from standard tree level results. Finally, we further study the infrared (IR) divergences of Carrollian amplitudes in massless scalar QED, gravity, and Yang-Mills theory. We show that Carrollian amplitudes in these theories naturally factorize, allowing us to provide an IR-safe definition for these objects.}

\maketitle
\flushbottom

\section{Introduction}


Scattering amplitudes describe the transition from asymptotic in-states to asymptotic out-states. While scattering amplitudes have traditionally been studied in the momentum eigenstate basis, in recent years alternative bases have gained a lot of attention, particularly in light of the two main proposals of flat space holography: celestial holography and Carrollian holography. With the aim of describing the S-matrix in $d-$dimensional flat space, celestial holography proposes that the dual celestial conformal field theory (CFT) lives on the $(d-2)-$dimensional celestial sphere at null infinity \cite{Strominger:2017zoo,Raclariu:2021zjz,Pasterski:2021raf,Donnay:2023mrd}, while Carrollian holography proposes that the dual theory is a ($d-1$)-dimensional Carrollian  CFT\footnote{ A Carroll manifold is a smooth manifold equipped with a degenerate (corank-1) `metric' and a vector field in its kernel. Carrollian field theories are field theories defined on such manifolds and can often be realized as `$c\rightarrow 0$ limits' \cite{Leblond,SenGupta:1966qer} of relativistic field theories~\cite{Bagchi:2016bcd, Bagchi:2019xfx}.}\cite{Donnay:2022aba,Bagchi:2022emh,Donnay:2022wvx,Bagchi:2025vri,Nguyen:2025zhg,Ruzziconi:2026bix}. 

Focusing on Carrollian holography in this work, we study Carrollian amplitudes, which are flat space scattering amplitudes written in position space at null infinity \cite{Donnay:2022aba,Bagchi:2022emh,Donnay:2022wvx,Bagchi:2025vri,Nguyen:2025zhg,Ruzziconi:2026bix,Mason:2023mti}. More concretely, Carrollian amplitudes are obtained as suitable integral transforms of momentum space scattering amplitudes. One popular choice is essentially a Fourier transform~\cite{Mason:2023mti}, although a class of `modified Mellin' transforms have also been considered in the literature~\cite{Bagchi:2022emh}. Part of the interest in Carrollian amplitudes stems from the fact that they have been shown to satisfy the Ward identities associated with global conformal Carrollian symmetries \cite{Donnay:2022aba,Bagchi:2022emh,Donnay:2022wvx,Bagchi:2025vri,Nguyen:2025zhg,Ruzziconi:2026bix,Mason:2023mti}. As a result, lower point Carrollian amplitudes functionally resemble correlation functions of a Carrollian CFT.

There are now multiple prescriptions \cite{Donnay:2022aba,Bagchi:2022emh,Donnay:2022wvx,Bagchi:2025vri,Nguyen:2025zhg,Ruzziconi:2026bix,Mason:2023mti} that build upon this kinematic statement and posit full-fledged \textit{dynamical} equivalences between bulk scattering amplitudes (in the Carrollian basis) and correlation functions of a putative dual Carrollian theory. 

Although solutions to the Carrollian conformal Ward identities have now been extracted from bulk computations in several different ways \cite{Donnay:2022aba,Bagchi:2022emh,Donnay:2022wvx,Bagchi:2025vri,Nguyen:2025zhg,Ruzziconi:2026bix,Mason:2023mti}, Carrollian theories have proven notoriously difficult to quantize so that explicit examples of candidate duals remain elusive. Progress in this regard has branched into a few main directions, one of which involves the `intrinsic' study of  properties of (quantum) Carrollian theories~\cite{deBoer:2023fnj, deBoer:2021jej, Baiguera:2022lsw, Cotler:2024xhb, Cotler:2025dau, Cotler:2025npu, Bagchi:2016bcd, Bagchi:2019xfx}. Another popular approach-- the `bottom-up' approach-- involves using the properties of bulk scattering amplitudes to constrain properties of a putative dual \cite{Donnay:2022aba,Bagchi:2022emh,Donnay:2022wvx,Bagchi:2025vri,Nguyen:2025zhg,Ruzziconi:2026bix,Mason:2023mti}. For this purpose, it is important to have a good number of exemplar Carrollian amplitudes at hand. Most of
the work on Carrollian amplitudes has focused on massless scattering in four dimensions at tree level, and there are only a few known examples at loop level \cite{Adamo:2024mqn,Liu:2024nfc,Long:2026rpq}.  In this work, we study various new examples of Carrollian loop amplitudes with the aim of understanding new analytic properties that emerge compared to their tree level counterparts.


Carrollian amplitudes-- much like their celestial counterparts~\cite{Arkani-Hamed:2020gyp}-- are famously `anti-Wilsonian'. The integral transforms used to construct these amplitudes usually involve integrating momentum space amplitudes over particle energies- such procedures could potentially source divergences from both the low energy and high energy regimes. Fortunately, the Fourier (like) integral kernels used in Carrollian prescriptions oscillate wildly at large energies. Moreover, one can also regulate the integrand by adding an infinitesimal exponential damping factor. These two facts are usually enough to remove `ultraviolet' divergences with the damping factor translating to an $i\epsilon$ prescription for the resulting Carrollian amplitude. In contrast, infrared (IR) divergences in massless theories often translate to divergences in corresponding Carrollian amplitudes at a given loop order- these must be handled with a little more care.

In parallel, the study of IR divergences in gauge theory and gravity has a long history and continues to be an active field \cite{Kinoshita:1962ur,Lee:1964is,Weinberg:1965nx,Agarwal:2021ais,Naculich:2011ry,Lippstreu:2025jit,Donnay:2026urd, Choi:2024mac, Choi:2024ajz,Donnay:2020lur,Agrawal:2023zea}. Amplitudes in massless scalar Quantum Electrodynamics (QED), Yang-Mills theory, and gravity were shown to factorize into a soft factor times a hard part. See e.g. \cite{Weinberg:1965nx,Agarwal:2021ais,Naculich:2011ry}. The soft factor takes an exponential form that contains all the IR divergences to all orders in perturbation theory. In \cite{Arkani-Hamed:2020gyp,Gonzalez:2021dxw,Magnea:2021fvy}, it was shown that by transforming scattering amplitudes to a boost eigenbasis, celestial amplitudes in gauge theory and gravity factorize into a conformally soft and a conformally hard part. In this work, we show that Carrollian amplitudes possess a similar property that leads to an IR-safe quantity.

As mentioned earlier, there are two main (related) prescriptions for obtaining Carrollian amplitudes from their momentum space counterparts-- the Fourier prescription and the modified Mellin prescription. Both prescriptions are consistent with the conformal Carrollian Ward identities, with the former arising as a specific instance of the latter. While previous computations have often adopted the Fourier prescription, we find that the modified Mellin prescription is well suited to  studying certain aspects of Carrollian loop amplitudes as well as their IR-safe generalizations. For this reason, we shall largely adopt this prescription in this work and emphasise its importance for our calculations.

The paper is organized as follows. In section \ref{sec:2}, we review some basic definitions of Carrollian amplitudes. In section \ref{sec:3}, we analyze various examples of Carrollian loop amplitudes and study in detail their properties. This includes rational four-point one-loop amplitudes in Yang-Mills theory, four-point MHV amplitudes in $\mathcal{N}=4$ super Yang-Mills and $\mathcal{N}=8$ supergravity at one-loop. We generalize the one-loop result of Yang-Mills to all loop orders by the  Bern-Dixon-Smirnov (BDS) formula \cite{Bern:2005iz}, obtaining an all loop expression for Carrollian amplitudes. We also consider the four-point amplitude for scalar exchange mediated by gravitons in the eikonal regime. We show that the corresponding Carrollian amplitudes exhibit logarithmic behaviors in the `Carroll time' $u$. We also compute the discontinuities of these Carrollian amplitudes up to $\mathcal{O}(G^3)$ (where $G$ is Newton's constant) and show that they are descendants of Carrollian Born amplitudes. We also compute the Carrollian amplitudes
of one-loop scalar box diagrams.
In section \ref{sec:4}, we examine the imprints of IR divergences on Carrollian amplitudes in massless scalar QED, gravity, and Yang-Mills theory. The IR factorization implies that Carrollian amplitudes in these theories factorize into  soft and a `modified' amplitude. While the soft part is time independent and contains all the IR divergences, the modified factor is time dependent and provides an IR-safe definition of Carrollian amplitudes. In section \ref{sec:5}, we conclude with some future directions.

\section{Carrollian Amplitudes: Preliminaries} \label{sec:2}

In this section, we review some basics of Carrollian amplitudes. In the current literature, there are two slightly different definitions of Carrollian amplitudes, one based on Fourier transforms, the other one based on modified Mellin transforms. We shall introduce both of them and set up the notation.

We begin with the planar Bondi coordinates $(u, r, z, \bar{z})$ of four dimensional Minkowski space used in \cite{Alday:2024yyj,He:2019jjk,Gonzo:2020xza},
\begin{equation}
    X^{\mu} = u \partial_z\partial_{\bar{z}}q^{\mu}(z,\bar{z}) +r q^{\mu}(z,\bar{z}) \, ,
\end{equation}
where the null vector
\begin{equation}
    q^{\mu}(z,\bar{z}) = \frac{1}{2} (1+|z|^2,z+\bar{z}, -i(z-\bar{z}), 1-|z|^2) \,,
\end{equation}
is parameterized by a point on the celestial sphere. $u,r\in \mathbb{R}$, $z\in\mathbb{C}$. The Minkowski metric expressed in the coordinate system is
\begin{equation}
    ds^2 = du dr - r^2 dzd\bar{z} \, ,
\end{equation}
where we used $(+,-,-,-)$ for the signature.
The planar Bondi coordinates can cover both past and future null infinity $\mathcal{I}^{\pm}$ in the conformal compactification. In particular, the past null infinity is reached by $r\rightarrow-\infty$ while the future null infinity is reached by $r\rightarrow+\infty$. In addition to covering both past and future null infinity, the other advantage of these planar Bondi coordinates is that they emerge naturally from the flat limit of Bondi coordinates on AdS as illustrated in \cite{Alday:2024yyj,Surubaru:2025fmg,Kulkarni:2025qcx,Adamo:2025bfr}.

At $\mathcal{I}^{\pm}$, the boundary metric becomes degenerate
\begin{equation}
    q_{ab}dx^a dx^b= 0 du^2- dz d\bar{z} \,,
\end{equation}
where $x^a=\{u,z,\overline{z}\}$. Together with a vector field $n^a\partial_a = \partial_u$ that is the kernel of the metric, 
\begin{equation}
    q_{ab}n^a = 0  \, ,
\end{equation}
they define a representative of the conformal Carrollian structure at the null boundary.

Working in these planar Bondi coordinates, Carrollian amplitudes 
are the scattering amplitudes of massless particles expressed in a basis where each external state
is associated with a point on the conformal boundary \cite{Bagchi:2022emh,Donnay:2022wvx,Mason:2023mti}. Carrollian amplitudes are naturally related to the flat limit of AdS correlators in position space \cite{Alday:2024yyj,Bagchi:2023cen,Surubaru:2025fmg,Kulkarni:2025qcx}. Most of the study of Carrollian amplitudes is focused on tree level  amplitudes. It is important to study Carrollian amplitudes at loop level, which is necessary for a self consistent putative dual theory. Therefore, we would like to provide more examples of Carrollian amplitudes at loop level with further study of their properties. 


To compute Carrollian amplitudes, one starts with the following parameterization of a general null momentum $p^\mu$:
\begin{equation}
    p^\mu = \epsilon\omega q^\mu = \frac{1}{2} \epsilon\,\omega(1+|z|^2,z+\bar{z}, -i(z-\bar{z}), 1-|z|^2) \, , \label{eq:pmupara}
\end{equation}
where $\omega$ is the light-cone energy, and $z$ and $\bar{z}$ are coordinates on the celestial sphere. $\epsilon = \pm 1$ is associated with outgoing ($+1$) and incoming $(-1)$ particles. The Lorentz invariant spinor-helicity variables can be written as
\begin{equation}
    \langle ij \rangle = \epsilon_i \epsilon_j\sqrt{\omega_i\omega_j}z_{ij} \, , \quad [ij] = -\sqrt{\omega_i\omega_j}\bar{z}_{ij} \, .
\end{equation}
The scalar products of null momenta are given by
\begin{equation}
    s_{ij} = 2p_i\cdot p_j = \epsilon_i\epsilon_j\omega_i\omega_j z_{ij}\bar{z}_{ij} \, .
\end{equation}
We introduce the first definition of Carrollian amplitudes via Fourier transforms \cite{Donnay:2022wvx,Mason:2023mti} with respect to light-cone energies,
\begin{equation}
\begin{split}
 C_n(&\{u_1,z_1,\bar{z}_1\}^{\epsilon_1}_{J_1},\dots,\{u_n,z_n,\bar{z}_n\}^{\epsilon_n}_{J_n} ) \\
 &=\prod_{i=1}^n\left( \int_0^{+\infty} \frac{d\omega_i}{2\pi} e^{i\epsilon_i \omega_i u_i} \right) A_n(\{\omega_1,z_1\bar{z}_1\}^{\epsilon_1}_{J_1},\dots, \{\omega_n, z_n,\bar{z}_n\}^{\epsilon_n}_{J_n}) \, ,
\end{split} \label{eq:Carrolliandef1}
\end{equation}
where $A_n$ are the scattering amplitudes in a momentum eigenstate basis.  $J$ denotes the particle helicity. It has been shown that Carrollian amplitudes can be interpreted as correlators in a putative Carrollian CFT at null infinities
\begin{equation}
     C_n(\{u_1,z_1,\bar{z}_1\}^{\epsilon_1}_{J_1},\dots,\{u_n,z_n,\bar{z}_n\}^{\epsilon_n}_{J_n} )=\langle \Phi^{\epsilon_1}_{J_1}(u_1,z_1,\bar{z}_1) \ldots \Phi^{\epsilon_n}_{J_n}(u_n,z_n,\bar{z}_n) \rangle \, .
\label{holographic identification}
\end{equation}
where $\Phi^{\epsilon_i}_{J_i}(u_i,z_i,\bar{z}_i)$ are conformal Carrollian primaries with weights 
\begin{equation}
    (k_i,\bar{k}_i) = \left(\frac{1+\epsilon_i J_i}{2},\frac{1-\epsilon_i J_i}{2} \right).
\label{carrollian weights}
\end{equation}

The other commonly used definition of Carrollian amplitudes is to perform modified Mellin transforms \cite{Banerjee:2018gce,Banerjee:2019prz,Banerjee:2020kaa, Bagchi:2022emh,Nguyen:2023miw} rather than Fourier transforms. We denote it by $\widetilde{C}$ as follows,
\begin{equation}
    \begin{split}
 \widetilde{C}_n(&\{\Delta_1, u_1,z_1,\bar{z}_1\}^{\epsilon_1}_{J_1},\dots,\{\Delta_n, u_n,z_n,\bar{z}_n\}^{\epsilon_n}_{J_n} ) \\
 &=\prod_{i=1}^n\left( \int_0^{+\infty} \frac{d\omega_i}{2\pi} e^{i\epsilon_i \omega_i u_i} \omega_i^{\Delta_i-1} \right) A_n(\{\omega_1,z_1\bar{z}_1\}^{\epsilon_1}_{J_1},\dots, \{\omega_n, z_n,\bar{z}_n\}^{\epsilon_n}_{J_n}) \, , \label{eq:Carrolliandef2}
\end{split}
\end{equation}
where $\Delta_i$ are the conformal dimensions of the Carrollian primaries. When $\Delta_i=1$, the two definitions of Carrollian amplitudes coincide. When $\Delta_i$ takes values of integers $\Delta_i>1$, the two definitions are related by $\partial_u$ derivatives,
\begin{equation}
\begin{split}
    &\widetilde{C}_n(\{\Delta_1=m_1+1, u_1,z_1,\bar{z}_1\}^{\epsilon_1}_{J_1},\dots,\{\Delta_n=m_n+1, u_n,z_n,\bar{z}_n\}^{\epsilon_n}_{J_n} )\\
    =&\left(\prod_{i=1}^n\left(-i\epsilon_i\right)^{m_i}\right)\partial_{u_1}^{m_1}\dots\partial_{u_n}^{m_n} C_n(\{u_1,z_1,\bar{z}_1\}^{\epsilon_1}_{J_1},\dots,\{u_n,z_n,\bar{z}_n\}^{\epsilon_n}_{J_n} ) \, .
\end{split}
\end{equation}
The advantage of Carrollian amplitudes is that it renders amplitudes in Yang-Mills and gravity finite due to the oscillating exponential factor of the time coordinate \cite{Banerjee:2019prz,Banerjee:2020kaa, Bagchi:2022emh,Nguyen:2023miw,Mason:2023mti,Stieberger:2024shv}. Throughout this work, we refer to both definitions, (\ref{eq:Carrolliandef1}) and (\ref{eq:Carrolliandef2}), as Carrollian amplitudes. 




\section{Carrollian loop Amplitudes} \label{sec:3}
In this section, we focus on four-point Carrollian amplitudes with external massless particles at loop level. After introducing four-point kinematics in four dimensions, we compute various examples of Carrollian loop amplitudes in Yang-Mills and gravity. We also compute the Carrollian amplitudes associated with scalar box diagrams at one-loop. We study in detail Carrollian eikonal amplitudes, and show that the discontinuities of Carrollian eikonal amplitude exhibit a simple structure and can be related to the descendants of Carrollian amplitudes at tree level.

\subsection{Finite One-loop Amplitudes in Yang-Mills Theory}
In this subsection, we compute Carrollian amplitudes associated with finite one-loop four-point amplitudes in Yang-Mills theory. An analogous study on celestial amplitudes has been carried out in \cite{Gonzalez:2020tpi,Albayrak:2020saa}. We shall see the similarities and differences between the Carrollian and celestial counterparts.

To compute four-point Carrollian amplitudes, we begin by rewriting the momentum conserving $\delta$-functions using the parameterization of the null momenta (\ref{eq:pmupara}). Without loss of generality, we have considered the scattering channel: $(12\rightarrow34)$. We then have
\begin{align}
\delta^{(4)}(\omega_1 q_1 +\omega_2 q_2 &-\omega_3 q_3-\omega_4 q_4) =  \frac{4}{\omega_4 |z_{14}|^2 |z_{23}|^2} \nonumber\\
&\times \delta\left( \omega_1 -\frac{z_{24} \bar{z}_{34}}{z_{12} \bar{z}_{13}} \omega_4 \right) \delta\left( \omega_2 -\frac{z_{14} \bar{z}_{34}}{z_{12}\bar{z}_{32}} \omega_4\right) \delta\left(\omega_3 + \frac{z_{24}\bar{z}_{14}}{z_{23}\bar{z}_{13}}\omega_4\right) \delta(r-\bar{r}) \, , \label{eq:4deltas}
\end{align}
where $r$ is the two-dimensional conformally invariant cross ratio:
\begin{equation}
   r = \frac{z_{12}z_{34}}{z_{23}z_{41}} \, . 
\end{equation}
 The $\delta$-function $\delta(r-\bar{r})$ imposes a reality condition on the cross ratio. From the bulk perspective, this reality condition is a reflection of the fact that $2-2$ scattering is kinematically constrained to occur on a plane.
 
Our choice of scattering channel constrains $r$ to be greater than $1$. The remaining two cross ratios are not independent of $r$ and instead can be expressed in terms of $r$ as follows
\begin{equation}
    r-1 = \frac{z_{13}z_{24}}{z_{32}z_{14}} \,  , \quad  \frac{r}{r-1} = \frac{z_{12}z_{34}}{z_{13}z_{24}} = z \, .
\end{equation}

For later use, we shall also express the Mandelstam variables in terms of celestial variables using the four-point kinematic constraints imposed by (\ref{eq:4deltas}). One finds
\begin{align}
s &=s_{12}=  (r-1) \frac{|z_{14}|^2 |z_{34}|^2}{|z_{13}|^2} \omega_4^2 \, , \label{eq:Mands} \\
u&=-s_{23}=  \frac{1-r}{r} \frac{|z_{14}|^2 |z_{34}|^2}{|z_{13}|^2} \omega_4^2 \, ,\label{eq:Mandu}\\
t=-s-u &=  -\frac{(r-1)^2}{r} \frac{|z_{14}|^2 |z_{34}|^2}{|z_{13}|^2} \omega_4^2 \, .\label{eq:Mands+u}
\end{align}
We consider several simple examples at one-loop in Yang-Mills theory. It is well known that all plus/minus helicity amplitudes vanish at tree level \cite{Parke:1986gb}. At one-loop the finite momentum space color-ordered amplitudes are written as rational functions of spinor helicity variables as \cite{Bern:1991aq},
\begin{equation}
    A_{\mathrm{YM}}(1^{\pm}, 2^{\pm}, 3^{\pm}, 4^{\pm}) = g^4 \left( \frac{[12][34]}{\langle 12\rangle \langle 34\rangle}\right)^{\pm 1}  = g^4 \left( \frac{\bar{z}_{12}\bar{z}_{34}}{z_{12}z_{34}}\right)^{\pm 1} \, .
\end{equation}
We compute the corresponding Carrollian amplitude defined by Fourier transforms (\ref{eq:Carrolliandef1}),
\begin{equation}
    \begin{split}
       & C_{\mathrm{YM}}(+,+,+,+) = g^4\int \prod_{i=1}^4\frac{d\omega_i}{2\pi} e^{-i\omega_1 u_1}e^{-i\omega_2 u_2} e^{i \omega_3 u_3} e^{i\omega_4 u_4} \frac{\bar{z}_{12}\bar{z}_{34}}{z_{12}z_{34}} \\
        &\times \frac{4}{\omega_4 |z_{14}|^2 |z_{23}|^2} \delta\left( \omega_1 -\frac{z_{24} \bar{z}_{34}}{z_{12} \bar{z}_{13}} \omega_4 \right) \delta\left( \omega_2 -\frac{z_{14} \bar{z}_{34}}{z_{12}\bar{z}_{32}} \omega_4\right) \delta\left(\omega_3 + \frac{z_{24}\bar{z}_{14}}{z_{23}\bar{z}_{13}}\omega_4\right) \delta(r-\bar{r})\\
        =& \frac{4g^4}{(2\pi)^4} \frac{\bar{z}_{12}\bar{z}_{34}}{z_{12}z_{34}}\frac{\delta(r-\bar{r})}{|z_{14}|^2|z_{23}|^2} \int_0^\infty\frac{d\omega_4}{\omega_4} e^{i\omega_4 x_4} \, , \label{eq:YMallplus}
    \end{split}
\end{equation}
where $g$ is the Yang-Mills coupling constant.
In order to simplify the notation, we have defined
\begin{align}
x_4 &=  u_4- \frac{z_{24} \bar{z}_{34}}{z_{12} \bar{z}_{13}} u_1-\frac{z_{14} \bar{z}_{34}}{z_{12}\bar{z}_{32}} u_2-\frac{z_{24}\bar{z}_{14}}{z_{23}\bar{z}_{13}} u_3  \nonumber\\
&= u_4 - \frac{r}{r-1} \frac{|z_{24}|^2}{|z_{12}|^2} u_1 -\frac{1}{r} \frac{|z_{34}|^2}{|z_{23}|^2} u_2 + (r-1) \frac{|z_{14}|^2}{|z_{13}|^2}u_3 \, . \label{eq:defx4}
\end{align}
The integral in (\ref{eq:YMallplus}) is implicitly regularized in the UV (i.e. at large $\omega$) by introducing an infinitesimal factor of $e^{-\epsilon\omega_4}$ to the integrand. See, e.g. \cite{Mason:2023mti} for detailed discussions. Furthermore, the integral (\ref{eq:YMallplus}) is also IR divergent, which shares the same property as the Carrollian gluon amplitudes at tree level \cite{Mason:2023mti}. One way to regularize it is to compute the $\partial_u$ descendant, e.g. 
\begin{equation}
\begin{split}
    \partial_{u_4}C_{\mathrm{YM}}(+,+,+,+) &= \frac{4i g^4}{(2\pi)^4} \frac{\bar{z}_{12}\bar{z}_{34}}{z_{12}z_{34}} \frac{\delta(r-\bar{r})}{|z_{14}|^2 |z_{23}|^2} \int_0^\infty d\omega_4 e^{i \omega_4 x_4} \\
    &=-\frac{4g^4}{(2\pi)^4} \frac{\bar{z}_{12}\bar{z}_{34}}{z_{12}z_{34}} \frac{\delta(r-\bar{r})}{|z_{14}|^2 |z_{23}|^2} \frac{1}{x_4} \, .
\end{split}
\end{equation}
The other way of regularization is to use the alternative definition of Carrollian amplitudes defined via modified Mellin transforms (\ref{eq:Carrolliandef2}),
\begin{equation}
    \begin{split}
        &\widetilde{C}_{\mathrm{YM}}(+,+,+,+) = g^4\int \prod_{i=1}^4\frac{d\omega_i}{2\pi} \omega_i^{\Delta_i-1} \, e^{-i\omega_1 u_1}e^{-i\omega_2 u_2} e^{i \omega_3 u_3} e^{i\omega_4 u_4} \frac{\bar{z}_{12}\bar{z}_{34}}{z_{12}z_{34}} \\
        &\times \frac{4}{\omega_4 |z_{14}|^2 |z_{23}|^2} \delta\left( \omega_1 -\frac{z_{24} \bar{z}_{34}}{z_{12} \bar{z}_{13}} \omega_4 \right) \delta\left( \omega_2 -\frac{z_{14} \bar{z}_{34}}{z_{12}\bar{z}_{32}} \omega_4\right) \delta\left(\omega_3 + \frac{z_{24}\bar{z}_{14}}{z_{23}\bar{z}_{13}}\omega_4\right) \delta(r-\bar{r})\\
        =& \frac{4g^4}{(2\pi)^4} \frac{\bar{z}_{12}\bar{z}_{34}}{z_{12}z_{34}} \frac{\delta(r-\bar{r})}{|z_{14}|^2 |z_{23}|^2} \left(\frac{z_{24} \bar{z}_{34}}{z_{12} \bar{z}_{13}}\right)^{\Delta_1-1} \left( \frac{z_{14} \bar{z}_{34}}{z_{12}\bar{z}_{32}}\right)^{\Delta_2-1} \left(\frac{z_{24}\bar{z}_{14}}{z_{23}\bar{z}_{31}} \right)^{\Delta_3-1} \int_0^\infty d\omega_4 \omega_4^{\sum_i \Delta_i -5} e^{i\omega_4 x_4} \\
        =&\frac{4g^4}{(2\pi)^4} \frac{\bar{z}_{12}\bar{z}_{34}}{z_{12}z_{34}} \frac{\delta(r-\bar{r})}{|z_{14}|^2 |z_{23}|^2} \left(\frac{z_{24} \bar{z}_{34}}{z_{12} \bar{z}_{13}}\right)^{\Delta_1-1} \left( \frac{z_{14} \bar{z}_{34}}{z_{12}\bar{z}_{32}}\right)^{\Delta_2-1} \left(\frac{z_{24}\bar{z}_{14}}{z_{23}\bar{z}_{31}} \right)^{\Delta_3-1} \frac{\Gamma(\Delta)}{(-i x_4)^{\Delta}} \, , \label{eq:allplusCt}
    \end{split}
\end{equation}
where we have denoted
\begin{equation}
    \Delta= \Delta_1+\Delta_2+\Delta_3+\Delta_4-4.
\end{equation}
We can express the Carrollian amplitude (\ref{eq:allplusCt}) in terms of a $SL(2,C)$ conformally invariant quantity $\mathcal{F}$ defined as follows
\begin{equation}
\begin{split}
        \mathcal{F}_{z=\bar{z}}&= \frac{|\frac{r}{r-1}|^2 |z_{23}|^2}{|z_{34}|^2 |z_{24}|^2} \left( u_4 -\frac{r}{r-1} \frac{|z_{24}|^2}{|z_{12}|^2} u_1 -\frac{1}{r} \frac{|z_{34}|^2}{|z_{23}|^2} u_2 + (r-1) \frac{|z_{14}|^2}{|z_{13}|^2}u_3 \right)^2  \\
        &=\frac{|\frac{r}{r-1}|^2 |z_{23}|^2}{|z_{34}|^2 |z_{24}|^2}  x_4^2 \, .
\end{split}
\end{equation}
As a  result, (\ref{eq:allplusCt}) can be written as
\begin{equation}
    \begin{split}
         &\widetilde{C}_{\mathrm{YM}}(+,+,+,+)\\
         =& \frac{4g^4}{(2\pi)^4} \frac{\bar{z}_{12}\bar{z}_{34}}{z_{12}z_{34}} \frac{\delta(r-\bar{r})}{|z_{14}|^2 |z_{23}|^2} \left(\frac{z_{24} \bar{z}_{34}}{z_{12} \bar{z}_{13}}\right)^{\Delta_1-1} \left( \frac{z_{14} \bar{z}_{34}}{z_{12}\bar{z}_{32}}\right)^{\Delta_2-1} \left(\frac{z_{24}\bar{z}_{14}}{z_{23}\bar{z}_{31}} \right)^{\Delta_3-1} \\
         &\times\left( \frac{|\frac{r}{r-1}|^2 |z_{23}|^2}{|z_{34}|^2 |z_{24}|^2}\right)^{\frac{\Delta}{2}} \frac{\Gamma(\Delta)}{(-i \mathcal{F})^{\frac{\Delta}{2}}} \, .
    \end{split}
\end{equation}
Note that the structure of this Carrollian amplitude is largely fixed by momentum conservation and the analytic structure of the momentum space amplitude. As mentioned earlier, momentum conservation invariably leads to the contact term in cross-ratios . This feature will persist at any loop order at four points.  On the other hand, the $u$ dependence is entirely fixed by the meromorphic nature of the momentum space amplitudes and the dimensions $\Delta_1$ through $\Delta_4$. 

Note that the singularities of the amplitude are far more intricate than the usual singularities encountered in Euclidean/Lorentzian CFTs \cite{Rychkov:2016iqz,Duffin} due to the presence of the contact term as well as the non trivial zero locus of $\mathcal{F}$. Better understanding of the analytic structure of such Carrollian amplitudes has been an ongoing effort \cite{Ruzziconi:2026bix,Nguyen:2025zhg,Liu:2024nfc}, but this goes beyond the scope of our present work.

In Yang-Mills theory, one-minus three-plus helicity amplitudes also vanish at tree level in flat space with the (1,3) spacetime signature \cite{Parke:1986gb}. However, see \cite{Guevara:2026qzd} for a recent study on non-vanishing one-minus amplitudes in Klein
space. At one-loop, similar to all-plus amplitudes, finite one-loop amplitudes are also written as a simple rational function \cite{Bern:1991aq}. 
\begin{equation}
    A_{\mathrm{YM}}(1^-, 2^+, 3^+, 4^+) =  g^4 \frac{\langle 13 \rangle [31][24]^2}{[12]\langle 23 \rangle \langle 34 \rangle [41]} = g^4 \left( \frac{z_{13} \bar{z}_{31} \bar{z}_{24}^2}{\bar{z}_{12} z_{23} z_{34} \bar{z}_{41}}\right) \, .
\end{equation}
We obtain the corresponding Carrollian amplitude 
\begin{equation}
\begin{split}
    \partial_{u_4}C_{\mathrm{YM}}(-,+,+,+) &= \frac{4i g^4}{(2\pi)^4} \left( \frac{z_{13} \bar{z}_{31} \bar{z}_{24}^2}{\bar{z}_{12} z_{23} z_{34} \bar{z}_{41}}\right)  \frac{\delta(r-\bar{r})}{|z_{14}|^2 |z_{23}|^2} \int_0^\infty d\omega_4 e^{i \omega_4 x_4} \\
    &=-\frac{4g^4}{(2\pi)^4} \left( \frac{z_{13} \bar{z}_{31} \bar{z}_{24}^2}{\bar{z}_{12} z_{23} z_{34} \bar{z}_{41}}\right) \frac{\delta(r-\bar{r})}{|z_{14}|^2 |z_{23}|^2} \frac{1}{x_4} \, ,
\end{split}
\end{equation}
and 
\begin{equation}
    \begin{split}
         &\widetilde{C}_{\mathrm{YM}}(-,+,+,+)\\
         =& \frac{4g^4}{(2\pi)^4} \left( \frac{z_{13} \bar{z}_{31} \bar{z}_{24}^2}{\bar{z}_{12} z_{23} z_{34} \bar{z}_{41}}\right)\frac{\delta(r-\bar{r})}{|z_{14}|^2 |z_{23}|^2} \left(\frac{z_{24} \bar{z}_{34}}{z_{12} \bar{z}_{13}}\right)^{\Delta_1-1} \left( \frac{z_{14} \bar{z}_{34}}{z_{12}\bar{z}_{32}}\right)^{\Delta_2-1} \left(\frac{z_{24}\bar{z}_{14}}{z_{23}\bar{z}_{31}} \right)^{\Delta_3-1} \\
         &\times\left( \frac{|\frac{r}{r-1}|^2 |z_{23}|^2}{|z_{34}|^2 |z_{24}|^2}\right)^{\frac{\Delta}{2}} \frac{\Gamma(\Delta)}{(-i \mathcal{F})^{\frac{\Delta}{2}}} \, .
    \end{split}
\end{equation}
Perhaps not unexpectedly, we have shown that finite one-loop amplitudes in Yang-Mills theory give rise to Carrollian amplitudes that have an analytic structure similar to the ones at tree level. 

\subsection{Carrollian MHV Loop Amplitudes in \texorpdfstring{$\mathcal{N}=4$}{N=4} Super Yang-Mills}
In this subsection, we consider a more interesting case, MHV amplitudes in $\mathcal{N}=4$ super Yang-Mills theory (SYM). The four-point MHV amplitudes in $\mathcal{N}=4$ SYM are known to all loop orders in the planar limit and are described by the famous Bern, Dixon and Smirnov (BDS) formula \cite{Bern:2005iz}. We will start with the one-loop case and show how the BDS formula relates Carrollian MHV amplitudes at the all-loops level to the ones at tree level. Similar to the previous subsection, the celestial analog of our result was computed in~\cite{Gonzalez:2020tpi}.

At one-loop, the dimensional regularization
method known as four-dimensional helicity (FDH) scheme \cite{Bern:2002zk} would be used. The momenta of the internal
particles are in $d=4-2\epsilon$ dimensions while all polarization vectors and the momenta of the external particle remain in $d=4$ dimensions. The four-point MHV amplitude at one-loop is expressed as 
\begin{equation}
    A_{\mathrm{1-loop}} = a \mathcal{M}^{(1)}_{\epsilon} A_{\mathrm{tree}} \, , \label{eq:A1loopMHV}
\end{equation}
where $A_{\mathrm{tree}}$ is given by the famous Parke-Taylor formula~\cite{Parke:1986gb} which, when recast in our kinematic variables, reads
\begin{equation}
    A_{\mathrm{tree}}(-,-,+,+) = g^2 \frac{\langle 12\rangle^3}{\langle 23 \rangle \langle 34\rangle \langle 41 \rangle} =g^2 r \frac{z_{12}\bar{z}_{34}}{\bar{z}_{12} z_{34}}. \label{eq:treeMHV}
\end{equation}
Additionally, $\mathcal{M}_{\epsilon}^{(1)}$ in~\eqref{eq:A1loopMHV} denotes the scalar box integral while $a$ denotes the 't Hooft coupling $a\equiv\frac{g^2 N}{8\pi^2} (4\pi e^{-\gamma_E})^\epsilon$ (where $\gamma_E$ is the Euler-Mascheroni constant).
The explicit form of $\mathcal{M}_{\epsilon}^{(1)}$ can be found in \cite{Gonzalez:2020tpi}, and goes as
\begin{equation}
\begin{split}
    \mathcal{M}_{\epsilon}^{(1)} =& -\frac{1}{\epsilon^2} e^{\epsilon \gamma_E}\gamma_{\Gamma} \frac{u}{\mu^2} \Bigg[ \left( \frac{\mu^2}{-u}\right)^{1+\epsilon} \, _2F_1( -\epsilon,1,1-\epsilon; 1+\frac{u}{s}) \\
    &- \left( \frac{\mu^2}{-s}\right)^{1+\epsilon} \,_2F_1(1,1,1-\epsilon; 1+\frac{u}{s}) \Bigg] \, , \label{eq:generic_box}
\end{split}
\end{equation}
where $\gamma_{\Gamma}=\Gamma(1+\epsilon)\Gamma^2(1-\epsilon)/\Gamma(1-2\epsilon)$, $\mu$ is  the dimensional regularization scale, and $_2 F_1$ denotes the ordinary hypergeometric function. For our purposes, it is also helpful to define the object $\mathcal{H}_1(r,\epsilon)$ to be 
\begin{equation}
\label{eq:H1Defn}
  \mathcal{H}_1(r,\epsilon) = \left( \frac{\mu^2}{-u}\right)^{-\epsilon} \mathcal{M}_{\epsilon}^{(1)} \, .
\end{equation}

By pulling an appropriate power of the cross ratio $u$ out from the square brackets, we see that the `non-trivial' portion of $\mathcal{M}_{\epsilon}^{(1)}$ depends entirely \textit{only} on the cross-ratios and not the energy $\omega_4$- this feature carries over to the Carrollian amplitude where the Carrollian time (i.e. $u$) dependence is largely unchanged and the complexity of the one-loop amplitude resides mostly in the $z$s via~\eqref{eq:generic_box}. Note also that the residue of the above $\epsilon$ double pole itself diverges as $u/s$ approaches $-1$- this is a signal of collinear divergences. We will not consider the kinematical regime in our present work.

To compute the corresponding Carrollian amplitude at one-loop, we follow the steps in \cite{Gonzalez:2020tpi}.  As we shall see, the implicit regulator in the modified Mellin transforms ensures that the UV contributions from the momentum space integrand are suppressed and do not lead to UV divergences in the Carrollian amplitude. On the other hand, the momentum space amplitude also suffers from divergences in the IR and these are reflected in the Carrollian amplitude. This is unlike the tree level case where the $\omega_4$ poles could be handled by an appropriate choice of scaling dimensions. The IR divergences at loop level stem from poles in the dimensional regulator $\epsilon$ and hence cannot be cured directly using the modified Mellin prescription.

To express (\ref{eq:A1loopMHV}) as a relation between the loop and tree Carrollian amplitudes, we first compute the modified Mellin transform of the tree level  MHV gluon amplitudes. The result can be found in \cite{Banerjee:2020vnt}. Here, we rewrite it with our conventions.
Starting from four-point tree level MHV amplitude (\ref{eq:treeMHV}),
we obtain the corresponding Carrollian amplitude
\begin{equation}
    \begin{split}
         &\widetilde{C}_{\mathrm{YM,tree}}(-,-,+,+)\\
         =& \frac{4g^2}{(2\pi)^4}  \frac{z_{12}\bar{z}_{34}}{\bar{z}_{12} z_{34}}\frac{r\delta(r-\bar{r})}{|z_{14}|^2 |z_{23}|^2} \left(\frac{z_{24} \bar{z}_{34}}{z_{12} \bar{z}_{13}}\right)^{\Delta_1-1} \left( \frac{z_{14} \bar{z}_{34}}{z_{12}\bar{z}_{32}}\right)^{\Delta_2-1} \left(\frac{z_{24}\bar{z}_{14}}{z_{23}\bar{z}_{31}} \right)^{\Delta_3-1} \\
         &\times\left( \frac{|\frac{r}{r-1}|^2 |z_{23}|^2}{|z_{34}|^2 |z_{24}|^2}\right)^{\frac{\Delta}{2}} \frac{\Gamma(\Delta)}{(-i \mathcal{F})^{\frac{\Delta}{2}}} \, .
    \end{split}
\end{equation}
Using (\ref{eq:A1loopMHV}), we find the modified Mellin transforms for MHV amplitude at one-loop
\begin{equation}
    \begin{split}
         &\widetilde{C}_{\mathrm{YM,one-loop}}(-,-,+,+)\\
         =& \frac{4\,a\, g^2}{(2\pi)^4}  \frac{z_{12}\bar{z}_{34}}{\bar{z}_{12} z_{34}}\frac{r\delta(r-\bar{r})}{|z_{14}|^2 |z_{23}|^2} \left(\frac{z_{24} \bar{z}_{34}}{z_{12} \bar{z}_{13}}\right)^{\Delta_1-1} \left( \frac{z_{14} \bar{z}_{34}}{z_{12}\bar{z}_{32}}\right)^{\Delta_2-1} \left(\frac{z_{24}\bar{z}_{14}}{z_{23}\bar{z}_{31}} \right)^{\Delta_3-1} \\
         &\times\frac{\mu^{2\epsilon}}{\left( \frac{r-1 |z_{14}|^2|z_{34}|^2}{r|z_{13}|^2}\right)^\epsilon} \mathcal{H}_1(r,\epsilon)\left( \frac{|\frac{r}{r-1}|^2 |z_{23}|^2}{|z_{34}|^2 |z_{24}|^2}\right)^{\frac{\Delta-2\epsilon}{2}} \frac{\Gamma(\Delta-2\epsilon)}{(-i \mathcal{F})^{\frac{\Delta-2\epsilon}{2}}}  \\
         =& \frac{a\mu^{2\epsilon}}{\left( \frac{r-1 |z_{14}|^2|z_{34}|^2}{r|z_{13}|^2}\right)^\epsilon} \mathcal{H}_1(r,\epsilon) e^{-2\epsilon \frac{\partial}{\partial \Delta}} \widetilde{C}_{\mathrm{YM,tree}}(-,-,+,+) \, ,
    \end{split} \label{eq:CYM1loopMHV}
\end{equation}
where $e^{-2\epsilon \frac{\partial}{\partial \Delta}}$ is a differential operator that shifts $\Delta =\Delta_1+\Delta_2+\Delta_3+\Delta_4-4$ by $-2\epsilon$. Note that the requisite change in scaling dimensions can naturally be incorporated into the modified Mellin prescription which is consequently a more convenient choice than the `fixed dimension' prescription of~\eqref{eq:Carrolliandef1}.

The Carrollian one-loop amplitude is written as differential operators acting on the tree level  one. The above expression is not manifestly symmetric in the conformal dimensions $\Delta$s. We can improve this by recasting it in terms of the shift operator defined in~\cite{Gonzalez:2020tpi}. This yields 

\begin{equation}
    \widetilde{C}_{\mathrm{one-loop}} = a \hat{\mathcal{M}}_\epsilon^{(1)} \widetilde{C}_{\mathrm{tree}} \, , \label{eq:C1loopfromCtree}
\end{equation}
where the operator $\hat{\mathcal{M}}_\epsilon^{(1)}$ is defined as
\begin{equation}
    \hat{\mathcal{M}}_\epsilon^{(1)} = \mathcal{H}_1(r,\epsilon) \hat{P}^\epsilon\,,
\end{equation}
with
\begin{equation}
    \hat{P} = \mu^2 r^{\frac{1}{3}}(r-1)^{1/3}\prod_{i<j}(z_{ij}\bar{z}_{ij})^{-\frac{1}{6}}\exp\left(-\frac{1}{2}\sum_{k=1}^4 \frac{\partial}{\partial \Delta_k} \right) \, . \label{eq:hatP}
\end{equation}
With this definition, it is straightforward to apply the BDS formula to Carrollian amplitudes, leading to an all-order expression for Carrollian amplitudes.

The all-loop four-point planar MHV amplitudes in momentum space is written as
\begin{equation}
    A_\mathrm{{all\, loops}} = \mathcal{M}_{\epsilon} A_{\mathrm{tree}} \, , \label{eq:Aallloop}
\end{equation}
where 
\begin{equation}
    \mathcal{M}_\epsilon = 1+\sum_{\ell=1}^\infty a^{\ell} \mathcal{M}^{(\ell)}_{\epsilon} \, .
\end{equation}
The $\mathcal{M}^{(\ell)}_\epsilon$ are functions of the cross ration $r$ and the regulator $\epsilon$ with a leading epsilon pole of order $2l$~\cite{Cachazo:2006mq}. Following our one-loop results, it is also useful to reexpress $\mathcal{M}^{(\ell)}_{\epsilon}$ in terms of $\mathcal{H}_\ell$, the $\ell$ loop analog of $\mathcal{H}_1$ (defined in~\eqref{eq:H1Defn}) as
\begin{equation}
\begin{split}
    \mathcal{M}^{(\ell)}_{\epsilon}&=\left( \frac{\mu^2}{-u}\right)^{\ell \epsilon} \mathcal{H}_\ell(r,\epsilon) \, ,
\end{split}
\end{equation}
The result~(\ref{eq:C1loopfromCtree}) for Carrollian one-loop amplitudes can now be generalized to higher loops as
\begin{equation}
    \widetilde{C}_{\ell\mathrm{-loop}} = a^\ell \hat{\mathcal{M}}^{(\ell)}_{\epsilon} \widetilde{C}_{\mathrm{tree}} \, ,
\end{equation}
where
\begin{equation}
    \hat{\mathcal{M}}^{(\ell)}_{\epsilon} = \mathcal{H}_\ell(r,\epsilon) \hat{P}^{\ell \epsilon} \, .
\end{equation}
We can re-sum the perturbative expansion to all loop orders,
\begin{equation}
    \widetilde{C}_{\mathrm{all\, loops}} = \hat{\mathcal{M}}_\epsilon \widetilde{C}_{\mathrm{tree}} \, ,
\end{equation}
with 
\begin{equation}
    \hat{\mathcal{M}}_{\epsilon} = 1+\sum_{\ell=1}^\infty a^\ell  \hat{\mathcal{M}}^{(\ell)}_{\epsilon} \, .
\end{equation}
The BDS formula implies that the prefactor in (\ref{eq:Aallloop}) has an iterative structure for the planar amplitude. As a result, it can be written in exponential form
\begin{equation}
    \mathcal{M}_\epsilon = 1+\sum_{\ell=1}^\infty a^{\ell} \mathcal{M}^{(\ell)}_{\epsilon} =\exp\left[ \sum_{\ell=1}^\infty a^\ell\left(f_\epsilon^{(\ell)} \mathcal{M}_{\ell \epsilon}^{(1)}+C^{(\ell)}+\mathcal{E}_\epsilon^{(\ell)}\right)\right] \, ,
\end{equation}
where $f_\epsilon^{(\ell)}$ are regular functions of $\epsilon$ and are related to the collinear and cusp anomalous dimensions. $C^\ell$ are numerical constants. $\mathcal{E}_\epsilon^{(l)}$ are of the order $\mathcal{O}(\epsilon)$.

One finds that the BDS formula leads to the Carrollian MHV amplitude to all loop orders
\begin{equation}
    \widetilde{C}_{\mathrm{all \, loops}} = \exp\left(\sum_{\ell=1}^\infty a^\ell\left(f_\epsilon^{(\ell)} \mathcal{H}_1(r,\ell\epsilon)+C^{(\ell)}+\mathcal{E}_\epsilon^{(\ell)}\right)\hat{P}^{\ell\epsilon}\right) \widetilde{C}_{\mathrm{tree}} \, .
\end{equation}
Much like in the celestial case, we note that the theme of loop amplitudes being obtained via appropriate shifting operators is not restricted to the BDS setup alone. Rather, this feature arises from two structures. First, factorization of the 1 loop-amplitude in momentum space into the tree amplitude and a `multiplicative' one-loop correction and, second, the relatively trivial (power law) dependence of the correction on the energies. These two structures are not uncommon, as we will see in some more examples going forward.

\subsection{Carrollian Loop Amplitudes in \texorpdfstring{$\mathcal{N}=8$}{N=8} Supergravity}
In the following subsections, we consider some examples of loop amplitudes in gravity. One of the simplest examples of such amplitudes is the one-loop four-point MHV amplitude in $\mathcal{N}=8$ supergravity. See, e.g. \cite{Kallosh:2008ru,Carrasco:2012ca}. The expression is given by
\begin{equation}
    M_{4, \,\mathcal{N}=8}^{(1)}=i
    \frac{\kappa^2}{2} s t u M_{\mathrm{tree}}(\mathcal{M}_{\epsilon}^{(1)}(s,t)+\mathcal{M}_{\epsilon}^{(1)}(s,u)+\mathcal{M}_{\epsilon}^{(1)}(t,u)) \, , \label{eq:M4gr1loop}
\end{equation}
where $\mathcal{M}_{\epsilon}^{(1)}$ is the massless scalar box diagram that appeared in one-loop four-point  MHV gluon amplitudes. 
$\kappa$ is the gravitational coupling constant. The tree level  four-point helicity amplitude of four gravitons is simply
\begin{equation}
    M_{\mathrm{tree}}(1^-,2^-,3^+,4^+) = \kappa^2\frac{\langle12\rangle^4[34]^4}{stu} \, .
\end{equation}
To compute the Carrollian amplitude associated with the tree amplitude, we express it as
\begin{equation}
    M_{\mathrm{tree}}(1^-,2^-,3^+,4^+)=\kappa^2\omega_4^2 \frac{|z_{14}|^2 |z_{34}|^2}{|z_{13}|^2} \left(r \frac{z_{12}\bar{z}_{34}}{\bar{z}_{12}z_{34}}\right)^2 \, .
\end{equation}
Performing the modified Mellin transforms, we find the corresponding Carrollian amplitude
\begin{equation}
\begin{split}
    &\widetilde{C}_{\mathrm{GR,tree}}(-,-,+,+) \\
    &=\frac{\kappa^2}{4\pi^4} \frac{ |z_{34}|^2}{|z_{13}|^2} \left(r \frac{z_{12}\bar{z}_{34}}{\bar{z}_{12}z_{34}}\right)^2 \frac{\delta(r-\bar{r})}{ |z_{23}|^2} \left(\frac{z_{24} \bar{z}_{34}}{z_{12} \bar{z}_{13}}\right)^{\Delta_1-1} \left( \frac{z_{14} \bar{z}_{34}}{z_{12}\bar{z}_{32}}\right)^{\Delta_2-1} \left(\frac{z_{24}\bar{z}_{14}}{z_{23}\bar{z}_{31}} \right)^{\Delta_3-1} \\
         &\times\left( \frac{|\frac{r}{r-1}|^2 |z_{23}|^2}{|z_{34}|^2 |z_{24}|^2}\right)^{\frac{\Delta+2}{2}} \frac{\Gamma(\Delta+2)}{(-i \mathcal{F})^{\frac{\Delta+2}{2}}} \, .
\end{split}
\end{equation}
Similar to the gluon case in (\ref{eq:C1loopfromCtree}), we compute the Carrollian amplitude at one-loop using (\ref{eq:M4gr1loop}).  We obtain
\begin{equation}
\begin{split}
    &\widetilde{C}_{\mathrm{GR,one-loop}}(-,-,+,+)\\
    =&a_G(\mathcal{F}_{st}(r,\epsilon)+\mathcal{F}_{su}(r,\epsilon)+\mathcal{F}_{tu}(r,\epsilon))\hat{P}^{\epsilon-1}\widetilde{C}_{\mathrm{GR,tree}}(-,-,+,+) \, . \label{eq:Ctgr1loop}
\end{split}
\end{equation}
where $a_G=i \kappa^2/2$. $\hat{P}$ is given by (\ref{eq:hatP}). The expressions of $\mathcal{F}$ can be read off from (\ref{eq:generic_box}). The Carrollian  one-loop amplitude (\ref{eq:Ctgr1loop}) shares the same property as the gluon case (\ref{eq:C1loopfromCtree}). It is also written as differential (weight-shifting) operators acting on the tree level  one.

\subsection{Carrollian Loop Amplitudes from Eikonal Amplitudes in Gravity}
 The next example of interest is the 't Hooft amplitude \cite{tHooft:1987vrq} that describes $2\to 2$ scattering of massless scalars with gravitational interaction in the eikonal limit defined as the regime where $s>>-t$ and $Gs\rightarrow\infty$ (where $G=\frac{\kappa^2}{8\pi}$ denotes Newton's constant). In the momentum basis, it is the same as the $1$ to $1$ amplitude in a shockwave background modulo the kinematic $\delta$ functions\footnote{For recent developments of gravitational eikonal, see \cite{DiVecchia:2023frv} for a comprehensive review.}. The formula for the 't Hooft amplitude is given by
 \begin{equation}\label{tHooftamp}
A_{\mathrm{eik}}=-\frac{2\pi i\,s}{\mu^2}\,\frac{\Gamma(1-i\,G\,s)}{\Gamma(i\,G\,s)}\left(\frac{4\,\mu^2}{-t}\right)^{1-i\,G\,s}\,,
\end{equation}
where $\mu$ is a mass scale acting as an IR regulator. This is equivalent to
\begin{equation}\label{tHooftamp2}
A_{\mathrm{eik}}=\underbrace{\frac{8\pi\,G\,s^2}{t}}_{\mathrm{Born}}\,\underbrace{\frac{\Gamma(-i\,G\,s)}{\Gamma(i\,G\,s)}\left(\frac{4\,\mu^2}{-t}\right)^{-i\,G\,s}}_{\mathrm{phase}}\,.
\end{equation}
The eikonal amplitude has a remarkable feature:  It is equal to the classical Born approximation times a phase which is all-orders in the coupling.

 In \cite{Adamo:2024mqn}, the $1$-to-$1$ Carrollian amplitude in the shockwave background was studied.  Here, we study the corresponding four-point Carrollian amplitude up to two loops where we observe some interesting properties. We begin by computing the Carrollian Born amplitude. Using the expressions of the Mandelstam variables in (\ref{eq:Mands})-(\ref{eq:Mands+u}), we find 
\begin{equation}
    \frac{s^2}{-t} = r \frac{|z_{14}|^2 |z_{34}|^2}{|z_{13}|^2} \omega_4^2  \, .
\end{equation}
 For just the eikonal example, we shall make use of the Fourier prescription~\eqref{eq:Carrolliandef1} and not the modified Mellin prescription~\eqref{eq:Carrolliandef2}. This is merely for simplicity since the Fourier prescription yields well defined Carrollian amplitudes for the eikonal example. Moreover, while the Mellin prescription of course yields different functional forms for the Carrollian amplitudes, the qualitative nature of our results will not be altered. Of course, the Fourier prescription can alternately be viewed as the Mellin prescriptions with all scaling dimensions $\Delta_i=1$. 
 
 Performing the Fourier transforms by the definition (\ref{eq:Carrolliandef1}), the Carrollian Born amplitude is simply
\begin{equation}
\begin{split}
        C_{\mathrm{Born}} =&-\frac{2G}{\pi^3} \frac{\delta(r-\bar{r})}{ |z_{23}|^2} r \frac{|z_{34}|^2}{|z_{13}|^2}  \frac{1}{x_4^2} \, .
\end{split} \label{eq:CBorn}
\end{equation}
To compute the Carrollian amplitude at one-loop from (\ref{tHooftamp}), we assume $Gs$ is small\footnote{The Fourier transform for Carrollian amplitudes contains a region of integration where this assumption breaks down. However, this region of integration will generically
receive quantum corrections as mentioned in \cite{Adamo:2024mqn}.} and use the standard expansion of the Gamma function near 0,
\begin{equation}
\Gamma(x)=\frac{1}{x}-\gamma_E+\frac{1}{2}\left(\frac{\pi^2}{6}+{\gamma_E^2}\right)x+\mathcal{O}(x^2) \, .
\end{equation}
Expanding (\ref{tHooftamp2}) to $O(G^2)$, we get
\begin{equation}
\begin{split}
     A_{\mathrm{eik}} =& \frac{8\pi G s^2}{-t} + \frac{8\pi G^2 s^2}{-t}\left[ 2i \gamma_E s- i s \log\left( \frac{4 \mu^2}{-t}\right) \right] + \mathcal{O}(G^3) \, , \label{eq:eikexpand}
\end{split}
\end{equation}
where $\gamma_E$ is the Euler-Mascheroni constant.
We rewrite the $O(G^2)$ terms inside the above square brackets as
\begin{equation}
    \begin{split}
       & 2i \gamma_E s- i s \log\left( \frac{4 \mu^2}{-t}\right)\\
       =&i(r-1) \frac{|z_{14}|^2 |z_{34}|^2}{|z_{13}|^2} \omega_4^2 \left[2 \gamma_E+2 \log(\omega_4) +\log\left( \frac{(r-1)^2 |z_{14}|^2 |z_{34}|^2}{4 r \mu^2 |z_{13}|^2}\right) \right] \, .
    \end{split}
\end{equation}
It is straightforward to compute the $\mathcal{O}(G^2)$ corrections to the Carrollian Born amplitude (\ref{eq:CBorn}),
\begin{equation}
    \begin{split}
        C_{\mathcal{O}(G^2)} =&\frac{2 G^2}{\pi^3} \frac{i r(r-1)\delta(r-\bar{r})}{|z_{14}|^2 |z_{23}|^2} \frac{|z_{14}|^4 |z_{34}|^4}{|z_{13}|^4} \int_0^\infty d\omega_4 \omega_4^3 e^{i\omega_4 x_4} \\
        &\times \left[2 \gamma_E+2 \log(\omega_4) +\log\left( \frac{(r-1)^2 |z_{14}|^2 |z_{34}|^2}{4 r \mu^2 |z_{13}|^2}\right) \right] \, , \label{eq:COG^2int}
    \end{split}
\end{equation}
which has the following two types of integrals that can be evaluated
\begin{subequations}
\begin{align}
    \int_0^\infty d\omega_4 \omega_4^3 e^{i\omega_4 x_4}&= \frac{6}{x_4^4} \, ,
    \\
    \int_0^\infty d\omega_4 \omega_4^3 \log(\omega_4) e^{i\omega_4 x_4}&= \frac{11-6 \gamma_E - 6 \log(-i x_4)}{x_4^4} \, .
    \end{align}
\end{subequations}
Therefore, the $\mathcal{O}(G^2)$ corrections to the Carrollian Born amplitude (\ref{eq:COG^2int}) become
\begin{equation}
    \begin{split}
        C_{\mathcal{O}(G^2)} =& \frac{2 G^2}{\pi^3} \frac{i r(r-1)\delta(r-\bar{r})}{|z_{14}|^2 |z_{23}|^2} \frac{|z_{14}|^4 |z_{34}|^4}{|z_{13}|^4} \\
        &\times \frac{22-12\log(-i x_4)+6\log\left( \frac{(r-1)^2 |z_{14}|^2 |z_{34}|^2}{4 r \mu^2 |z_{13}|^2}\right) }{x_4^4} \, , \label{eq:Ceikoneloop}
    \end{split}
\end{equation}
where the terms involving $\gamma_E$ cancel nicely. This is consistent with the calculations in \cite{Adamo:2024mqn} where $\gamma_E$ can be absorbed by the IR regulator. In addition, the $\log$ terms can be combined into a Lorentz invariant quantity $\log\left( \frac{(r-1)^2 |z_{14}|^2 |z_{34}|^2}{4 r \mu^2 |z_{13}|^2 x_4^2} \right)\, $.

The presence of the $\log$ terms in (\ref{eq:Ceikoneloop}) means that  the Carrollian amplitude at one-loop contains a discontinuity. We find that
\begin{equation}
    \text{Disc}(C_{\mathcal{O}(G^2)} )\sim \frac{(r-1)|z_{14}|^2 |z_{34}|^2}{|z_{13}|^2 x_4^2}C_{\mathrm{Born}} = |z_{12}|^2 \partial_{u_1}\partial_{u_2}C_{\mathrm{Born}} \, . \label{eq:discC}
\end{equation}
Interestingly, the discontinuity of the Carrollian amplitude at one-loop is a $\partial_u$ descendant of the Carrollian Born amplitude.

We would like to study the higher order corrections in $G$ of the Carrollian amplitude to see if the properties that we observed above are generic. Expanding the 't Hooft amplitude (\ref{tHooftamp}) to order of $G^3$,
\begin{equation}
\begin{split}
     A_{\mathrm{eik}} =& \frac{8\pi G s^2}{-t} + \frac{8\pi G^2 s^2}{-t}\left[ 2i \gamma_E s- i s \log\left( \frac{4 \mu^2}{-t}\right) \right] \\
     &+\frac{8\pi G^3 s^2}{-t}\left[-2\gamma_E^2 s^2+2\gamma_E s^2 \log\left( \frac{4 \mu^2}{-t}\right)-\frac{1}{2}s^2 \log^2\left( \frac{4 \mu^2}{-t}\right) \right]
     + \mathcal{O}(G^4) \, . \label{eq:eikexpandG3}
\end{split}
\end{equation}
We perform the Fourier transforms to compute the associated order $G^3$ corrections to the Carrollian Born amplitude. After a lengthy calculation, we obtain
\begin{equation}
    \begin{split}
        C_{\mathcal{O}(G^3)} =& \frac{8G^3}{\pi^3} \frac{r(r-1)^2}{|z_{14}|^2|z_{23}|^2} \frac{|z_{14}|^6 |z_{34}|^6}{|z_{13}|^6}\delta(r-\bar{r}) \frac{1}{x_4^6}\\
        &\times\Bigg[ 225+10 \pi^2 +15\left[-2 \log(-i x_4)+\log\left( \frac{(r-1)^2 |z_{14}|^2 |z_{34}|^2}{4 r \mu^2 |z_{13}|^2}\right) \right]^2+ \\
        &+ 137\left[-2 \log(-i x_4)+\log\left( \frac{(r-1)^2 |z_{14}|^2 |z_{34}|^2}{4 r \mu^2 |z_{13}|^2}\right) \right] \Bigg] \, .
    \end{split}
\end{equation}
All the terms involving $\gamma_E$ cancel again nicely. It turns out that  the double discontinuity of the Carrollian amplitude at two loops is related to the Carrollian Born amplitude,
\begin{equation}
    \text{Disc}(\text{Disc}(C_{\mathcal{O}(G^3)})) \sim \left(\frac{(r-1)|z_{14}|^2 |z_{34}|^2}{|z_{13}|^2 x_4^2} \right)^2C_{\mathrm{Born}} = |z_{12}|^4 \partial_{u_1}^2\partial_{u_2}^2C_{\mathrm{Born}}
\end{equation}
As we can see, it is also a $\partial_u$ descendant of the Carrollian Born amplitude similar to the formula at one-loop (\ref{eq:discC}). It would be interesting to perform calibrations to higher orders in $G$. We expect the properties that we found to continue to hold in higher orders.

\subsection{Carrollian Amplitudes of One-Loop Scalar Box Diagrams}
In this subsection, we evaluate the Carrollian amplitude for the four-point one-loop box diagram in massless $\phi^3$.  Practically, this diagram is important as it appeared in the one-loop amplitudes in gauge theory (\ref{eq:A1loopMHV}) and gravity (\ref{eq:M4gr1loop}). From a more theoretical standpoint, the Carrollian loop amplitudes we have studied so far have differed in analytic structure from their tree level counterparts only in the $u$ dependent portions of the amplitude. Of course, this is largely because the examples we had encountered earlier had rather simple dependences on the energy. In contrast, the one-loop box features a non-trivial energy dependence which will translate to new structures in both the $u$s and the $\Delta$s. For instance, we will find that the corresponding Carrollian amplitude will turn out to have logs in the $u$ variable. 

The expression of the box diagram can be found in \cite{Smirnov2012}.  It is written as
\begin{align}
   & \mathcal{A}(p_1,p_2,p_3,p_4)=\frac{i\pi^{\frac{d}{2}}}{su}\Biggl(\frac{4}{\epsilon^2}-\frac{2}{\epsilon}\left(\log\left(\frac{-s}{\mu^2}\right)+\log\left(\frac{-u}{\mu^2}\right)+2\gamma_E\right)\nonumber \\
    &+2\log\left(\frac{-s}{\mu^2}\right)\log\left(\frac{-u}{\mu^2}\right)+2\gamma_E\left(\log\left(\frac{-s}{\mu^2}\right)+\log\left(\frac{-u}{\mu^2}\right)\right)+4\gamma_E^2-\frac{4\pi^2}{3}+\mathcal{O}(\epsilon)\Biggr) \, ,
\end{align}
where we dimensionally regulate with the mass parameter $\mu$  and $d=4-2\epsilon$. We have expanded it in $\epsilon$ around 0.
This amplitude can be expressed in the form 
\begin{align}
    \mathcal{A}(p_1,p_2,p_3,p_4)&=
    \frac{1}{su}\left(\alpha+\beta\left(\log\left(\frac{-s}{\mu^2}\right)+\log\left(\frac{-u}{\mu^2}\right)\right)+\gamma\log\left(\frac{-s}{\mu^2}\right)\log\left(\frac{-u}{\mu^2}\right)\right),
\end{align}
where we have neglected terms from $\mathcal{O}(\epsilon)$ onward and have defined
\begin{subequations}
\begin{align}
    \alpha\equiv& i\pi^{\frac{d}{2}}\left(\frac{4}{\epsilon^2}-\frac{4\gamma_E}{\epsilon}+4\gamma_E^2-\frac{4\pi^2}{3}\right),
    \\
    \beta\equiv& i\pi^{\frac{d}{2}}\left(-\frac{2}{\epsilon}+2\gamma_E\right),
    \\
    \gamma\equiv&2i\pi^{\frac{d}{2}}.
\end{align}
\end{subequations}

Substituting (\ref{eq:Mands}) and (\ref{eq:Mandu}) above, we obtain the following.
\begin{align}
 & \mathcal{A}(p_1,p_2,p_3,p_4)=
    \frac{-1}{r\left(\frac{z_{24} \bar{z}_{34}}{z_{12} \bar{z}_{13}}\right)^2\omega^4_4\vert z_{14}\vert^4}\nonumber\\
    &\times\left(\alpha+\beta\left(\log\left(\frac{-r\left(\frac{z_{24} \bar{z}_{34}}{z_{12} \bar{z}_{13}}\right)\omega_4^2\vert z_{14}\vert^2}{\mu^2}\right)+\log\left(\frac{\left(\frac{z_{24} \bar{z}_{34}}{z_{12} \bar{z}_{13}}\right)\omega_4^2\vert z_{14}\vert^2}{\mu^2}\right)\right) \right.\nonumber \\
    &+\left. \gamma\log\left(\frac{-r\left(\frac{z_{24} \bar{z}_{34}}{z_{12} \bar{z}_{13}}\right)\omega_4^2\vert z_{14}\vert^2}{\mu^2}\right)\log\left(\frac{\left(\frac{z_{24} \bar{z}_{34}}{z_{12} \bar{z}_{13}}\right)\omega_4^2\vert z_{14}\vert^2}{\mu^2}\right)\right), \label{eq:A4boxstep2}
\end{align}

To evaluate the modified Mellin transform, it is sufficient to note the following results, which hold for real $p,q$ and $\text{Re}(a)>0$:
\begin{subequations}
\begin{align}
\int_0^\infty{dx e^{-ax}x^{\Delta-1}\ln(x)}&=\frac{1}{a^{\Delta}}\Gamma(\Delta)\left(\psi^{(0)}(\Delta)-\ln(a)\right) \, ,
\\
\int_0^\infty{dx e^{-ax}x^{\Delta-1}\ln(px)\ln(qx)}&=\frac{1}{a^{\Delta}}\Gamma(\Delta)\left(\left(\psi^{(0)}(\Delta)-\ln\left(\frac{a}{p}\right)\right)\left(\psi^{(0)}(\Delta)-\ln\left(\frac{a}{q}\right)\right)+\psi^{(1)}(
\Delta
)\right).
\end{align}
\end{subequations}
Here, $\Gamma(x)$ denotes the Gamma function and $\psi^{(n)}(x)$ is the order-$n$ polygamma function. In particular, $\psi^{(0)}(x)$ is the digamma function. Making use of these results, one can compute the Carrollian amplitude from (\ref{eq:A4boxstep2}). To make the evaluation clear, we will first present the result for the terms proportional to $\alpha$,$\beta$ and $\gamma$- We will denote these by $\widetilde{C}_\alpha$,$\widetilde{C}_\beta$,$\widetilde{C}_\gamma$ and then assemble them to obtain the main result. The full amplitude is given by 
\begin{equation}
\widetilde{C}_{\mathrm{box}}=\widetilde{C}_\alpha+\widetilde{C}_\beta+\widetilde{C}_\gamma\,,
\end{equation}
where we have

\begin{subequations}
\begin{align}
\widetilde{C}_\alpha&=-\frac{4\alpha}{r\vert z_{14}\vert^6 \vert z_{23}\vert^2}\left(\frac{z_{12}\overline{z}_{13}}{z_{24}\overline{z}_{34}}\right)^2\left(\left(\frac{z_{24}\overline{z}_{34}}{z_{12}\overline{z}_{13}}\right)^{\Delta_1-1}\left(\frac{z_{14}\overline{z}_{34}}{z_{12}\overline{z}_{32}}\right)^{\Delta_2-1}\left(\frac{z_{24}\overline{z}_{14}}{z_{23}\overline{z}_{31}}\right)^{\Delta_3-1}\right)
\left(\frac{\Gamma(\Delta-8)}{(ix_4)^{\Delta-8}}\right) \, ,
\\
\widetilde{C}_\beta&=-\frac{16\beta}{r\vert z_{14}\vert^6 \vert z_{23}\vert^2}\left(\frac{z_{12}\overline{z}_{13}}{z_{24}\overline{z}_{34}}\right)^2\left(\left(\frac{z_{24}\overline{z}_{34}}{z_{12}\overline{z}_{13}}\right)^{\Delta_1-1}\left(\frac{z_{14}\overline{z}_{34}}{z_{12}\overline{z}_{32}}\right)^{\Delta_2-1}\left(\frac{z_{24}\overline{z}_{14}}{z_{23}\overline{z}_{31}}\right)^{\Delta_3-1}\right)\left(\frac{\Gamma(\Delta-8)}{(ix_4)^{\Delta-8}}\right) \nonumber
\\
&\left(\psi^{(0)}(\Delta-8)+\ln\left(-\frac{r\vert z_{14}\vert^4\left(\frac{z_{24}\overline{z}_{34}}{z_{12}\overline{z}_{13}}\right)^2}{(i x_4)^4\mu^4}\right)\right) \, ,
\\
\widetilde{C}_\gamma&=-\frac{16\gamma}{r\vert z_{14}\vert^6 \vert z_{23}\vert^2}\left(\frac{z_{12}\overline{z}_{13}}{z_{24}\overline{z}_{34}}\right)^2\left(\left(\frac{z_{24}\overline{z}_{34}}{z_{12}\overline{z}_{13}}\right)^{\Delta_1-1}\left(\frac{z_{14}\overline{z}_{34}}{z_{12}\overline{z}_{32}}\right)^{\Delta_2-1}\left(\frac{z_{24}\overline{z}_{14}}{z_{23}\overline{z}_{31}}\right)^{\Delta_3-1}\right)\left(\frac{\Gamma(\Delta-8)}{(ix_4)^{\Delta-8}}\right) \nonumber
\\
&\left(\left(\psi^{(0)}(\Delta-8)-\frac{1}{2}\ln\left(-\frac{r\vert z_{14}\vert^2\left(\frac{z_{24}\overline{z}_{34}}{z_{12}\overline{z}_{13}}\right)}{(i x_4)^2\mu^2}\right)\right)\left(\psi^{(0)}(\Delta-8)-\frac{1}{2}\ln\left(\frac{\vert z_{14}\vert^2\left(\frac{z_{24}\overline{z}_{34}}{z_{12}\overline{z}_{13}}\right)}{(i x_4)^2\mu^2}\right)\right)+\psi^{(1)}(
\Delta-8
)\right) \, .
\end{align}
\end{subequations}
For this example, we define $\Delta=\Delta_1+\Delta_2+\Delta_3+\Delta_4$.
We therefore see the appearance of logarithms in $u$ for this Carrollian amplitude, as well as novel structural dependences on the scaling dimension $\Delta$. It would be interesting to better understand the analytic structures emerging here in the future, e.g. discontinuities of these Carrollian loop amplitudes.

\section{IR Divergences in Carrollian Amplitudes} \label{sec:4}

Gauge theory and gravity have a very rich infrared structure. There is a well-known exponentation of soft divergences \cite{Weinberg:1965nx,Agarwal:2021ais,Naculich:2011ry}. It has been shown in \cite{Arkani-Hamed:2020gyp,Gonzalez:2021dxw,Magnea:2021fvy} that, for celestial amplitudes, such exponentation leads to factorization into soft and hard parts. In this section, we explore the imprints of these exponentiations on Carrollian amplitudes.

\subsection{QED}

In this subsection, we consider massless scalar QED as an example to show how Carrollian amplitudes factorize. It is known that the IR divergences in this theory exponentiate \cite{Weinberg:1965nx}
\begin{equation}
    A=e^{B} A_0 \, ,
\end{equation}
where $A$ is a general $n$ particle amplitude (we suppress all labels and indices for brevity) and $A_0$ does not depend on $\Lambda_{\mathrm{IR}}$. The exponential factor is given by
\begin{equation}
     e^B = e^{-\alpha \sum_{i<j}Q_i Q_j \log|\frac{1}{2} p_i\cdot p_j|} \, ,
\end{equation}
where 
\begin{equation}
    \alpha = \frac{e^2}{4\pi^2} \log(\Lambda_{\mathrm{IR}}) \,, 
\end{equation}
contains the coupling and an IR cutoff scale $\Lambda_{\mathrm{IR}}$. $Q_i$ is the charge of each scalar.
We compute the corresponding Carrollian amplitudes by (\ref{eq:Carrolliandef1}). One can immediately see that they factorize into a soft part and a `modified' amplitude
\begin{equation}
\begin{split}
    C(\{u_i, z_i, \bar{z}_i\}) =& \int \prod_j \frac{d\omega_j}{2\pi} e^{i\epsilon_j \omega_j u_j} e^{-\alpha\sum_{k<l}Q_k Q_l \log|\frac{1}{2}  p_k\cdot p_l|} A_0 \\
    =&C_{\mathrm{soft}}C_{\mathrm{mod}} \, , \label{eq:QEDfactorized}
\end{split}
\end{equation}
where 
\begin{subequations}
\label{eq:ChardSoft}
\begin{align}
    C_{\mathrm{soft}} &= \prod_{i<j}\left( \frac{z_{ij}\bar{z}_{ij}}{4}\right)^{-\alpha Q_i Q_j} \, ,
\label{eq:CHard}
\\
  C_{\mathrm{mod}} &= \int \prod_j \frac{d\omega_j}{2\pi} e^{i\epsilon_j \omega_j u_j} \omega_j^{\alpha Q_j^2}A_0 \, .
    \end{align}
\end{subequations}
In (\ref{eq:QEDfactorized}), we use charge conservation to write $\prod_{i<j}(\omega_i\omega_j)^{-\alpha Q_iQ_j}=\prod_{i}\omega_i^{\alpha Q_i^2}$.
Notice that the $C_{\mathrm{mod}}$ part for the Carrollian amplitudes has operators with conformal dimensions $\Delta_i = 1+\alpha Q_i^2$.  


Again, we see that the modified Mellin prescription emerges somewhat naturally in this context. More precisely, although we started off with the Fourier prescription~\eqref{eq:Carrolliandef1} for the full amplitude $A$,~\eqref{eq:QEDfactorized} and~\eqref{eq:ChardSoft} tell us that this translates to using the modified Mellin prescription~\eqref{eq:Carrolliandef2} for $A_0$. Additionally, while the Fourier prescription does yield a splitting of the Carrollian amplitude into $C_{\mathrm{mod}}$ and $C_{\mathrm{mod}}$, the former still features formal divergences in scaling dimensions. Curiously, these could be negated by using the modified Mellin prescription from the get go. Specifically, it is easily seen that the modified Mellin analog of~\eqref{eq:QEDfactorized} reads
\begin{equation}
\begin{split}
    \widetilde{C}(\{u_i, z_i, \bar{z}_i, \Delta_i\}) =& \int \prod_j \frac{d\omega_j}{2\pi} e^{i\epsilon_j \omega_j u_j} e^{-\alpha\sum_{k<l}Q_k Q_l \log|\frac{1}{2}  p_k\cdot p_l|} A_0 \\
    =&\widetilde{C}_{\mathrm{soft}}\widetilde{C}_{\mathrm{mod}} \, , \label{eq:QEDfactorized2}
\end{split}
\end{equation}
with 
\begin{equation}
    \widetilde{C}_{\mathrm{soft}} = \prod_{i<j}\left( \frac{z_{ij}\bar{z}_{ij}}{4}\right)^{-\alpha Q_i Q_j} \, ,
\end{equation}
as before, but with the modified factor undergoing modifications as 
\begin{equation}
\label{eq:CHard2}
    \widetilde{C}_{\mathrm{mod}} = \int \prod_j \frac{d\omega_j}{2\pi} e^{i\epsilon_j \omega_j u_j} \omega_j^{\alpha Q_j^2+\Delta_j-1}A_0 \, .
\end{equation}
Consequently, it is possible to achieve a finite result for $\widetilde{C}_{\mathrm{mod}}$ if we `renormalize' the scaling dimensions $\Delta_j$s to soak up the $\alpha$ divergences. While this `renormalization' is at this stage merely a formal procedure to remove divergences, it would be interesting to search for deeper interpretations.

$\widetilde{C}_{\mathrm{mod}}$ contains all the dependence on $u$ while the soft part depends only on the coordinates of the celestial sphere $z$ and $\bar{z}$. Stripping of the soft part, we obtain an IR-safe definition of Carrollian amplitudes.


Formally, the stripping of the soft factors corresponds to considering scattering amplitudes not between momentum eigenstates, but rather between suitable coherent superpositions of such eigenstates, as done in (for instance) the Faddeev-Kulish prescription~\cite{Kulish:1970ut}. Again, these statements are not unique to the Carrollian approach and have received attention in previous celestial works ~\cite{Arkani-Hamed:2020gyp, Choi:2017bna, Choi:2017ylo}.

\subsection{Gravity}

In this subsection, we consider the scattering of gravitons in a general quantum theory of gravity.  It is known that the IR divergences for gravity amplitudes also exponentiate \cite{Weinberg:1965nx,Naculich:2011ry}
\begin{equation}
    A= e^B A_0 \, ,
\end{equation}
where, again, $A_0$ does not depend on the IR cutoff $\Lambda_{\mathrm{IR}}$ although UV finiteness may require the use of a UV cutoff $\Lambda_{\mathrm{UV}}$. Here, $B$ is defined to be
\begin{equation}
\begin{split}
     B=&-\gamma \sum_{i,j} (p_i\cdot p_j) \log(p_i\cdot p_j) \\
     &=-\frac{\gamma}{2} \sum_{i,j}\epsilon_i\epsilon_j \omega_i\omega_j |z_{ij}|^2 \log|z_{ij}|^2 \, .
\end{split}
\end{equation}
We have used momentum conservation to get to the second line. The constant $\gamma$ is defined as
\begin{equation}
    \gamma = \frac{G}{\pi}\log(\Lambda_{\mathrm{IR}}) \, .
\end{equation}
The corresponding Carrollian amplitudes also factorize as
\begin{equation}
\begin{split}
    C(u_i, z_i, \bar{z}_j) =&\int\prod_j \frac{d\omega_j}{2\pi} e^{i\epsilon_j \omega_j u_j} e^{-\frac{\gamma}{2} \sum_{i,j}\epsilon_i \epsilon_j \omega_i\omega_j |z_{ij}|^2 \log|z_{ij}|^2} A_0  \\
    =&C_{\mathrm{soft}}C_{\mathrm{mod}} \, , \label{eq:CsofthardQED}
\end{split}
\end{equation}
where 
\begin{subequations}
\begin{align}
    C_{\mathrm{soft}} &= e^{\frac{\gamma}{2}\sum_{i,j} |z_{ij}|^2 \log|z_{ij}|^2 \partial_{u_i}\partial_{u_j}} \, ,
\\
    C_{\mathrm{mod}} &= \int \prod_j \frac{d\omega_j}{2\pi} e^{i\epsilon_j \omega_j u_j} A_0 \,.
    \end{align}
\end{subequations}
Compared to the case of massless scalar QED (\ref{eq:CsofthardQED}), the main difference is that $C_{\mathrm{soft}}$ for gravity contains $\partial_u$ and is hence a differential operator rather than a standalone $c$-number contribution to the amplitude. It acts on $C_{\mathrm{mod}}$ to create $\partial_u$ descendants.  The hard part $C_{\mathrm{mod}}$ which is computed by the Fourier transforms of $A_0$ has operators with conformal dimension $\Delta_i=1$.  We can also utilize the modified Mellin prescription to obtain similar results.  Similar to the QED case, stripping of the soft part leads to an IR-safe definition of Carrollian amplitudes in gravity. 

Note that unlike the QED case, $C_{\mathrm{mod}}$ does not depend on $\gamma$ and, by extension, the IR cutoff. For this reason, we do not require further `renormalization' after stripping off the soft contribution $C_{\mathrm{soft}}$ and can land on finite amplitudes regardless of whether we work with the Fourier or modified Mellin prescriptions. This is quite unlike the situation in QED. It would be interesting to pursue a more thorough comparison of the two theories in this regard.

\subsection{Yang-Mills}
In this subsection, utilizing the known results of IR factorization of amplitudes in Yang-Mills theory\footnote{For the infrared structure of gauge theories, see e.g. \cite{Agarwal:2021ais} for a review.}, we show that the corresponding Carrollian amplitudes factorize nicely into a soft part and a part whose dependence on the IR cutoff can be `renormalized away' as before. A similar analysis on celestial amplitudes can be found in \cite{Gonzalez:2021dxw,Magnea:2021fvy}. 

It has been shown that an arbitrary $n$-particle amplitude $A_n$ (we will retain the $n$ subscript for just this subsection) in Yang-Mills theory factorizes into an IR divergent piece times an IR finite part as \cite{Sen:1982bt,Dixon:2008gr,Gardi:2009qi,Becher:2009qa,Feige:2014wja}
\begin{equation}
    {A}_n=\mathcal{Z}_n\left( \frac{p_i}{\mu}, \alpha_s(\mu), \epsilon\right) \mathcal{H}_n\left( \frac{p_i}{\mu}, \alpha_s(\mu), \epsilon\right) \, ,
\end{equation}
where dimensional regularization was used as $d=4-2\epsilon$ with $\epsilon<0$. While $\mathcal{H}_n$ is finite when $\epsilon \rightarrow0$, $\mathcal{Z}_n$ contains all infrared poles and satisfies a renormalization group equation, which can be solved in an exponential form
\begin{equation}
    \mathcal{Z}_n\left( \frac{p_i}{\mu}, \alpha_s(\mu), \epsilon\right)= P\exp\left[ \frac{1}{2}\int_0^{\mu^2} \frac{d\lambda^2}{\lambda^2}\Gamma_n\left( \frac{p_i}{\lambda}, \alpha_s(\lambda),\epsilon\right)\right] \, . \label{eq:Zndef}
\end{equation}
$\Gamma_n$ is called the soft anomalous dimension matrix. The symbol $P$ denotes the path ordering of these matrices according to the ordering of the
scales. $\alpha_s(\lambda)$ is the $d$ dimensional running coupling that satisfies the renormalization group equation. Note that $\Gamma_n$ is finite when $\epsilon \rightarrow0$. All the IR singularities come from the scale integral.

Color conservation strongly constrains the form of the soft anomalous dimension matrix. Specifically, one finds that the constraints allow us to recast the matrix as the sum of the so-called color-dipole $\Gamma_n^{\text{dipole}}$ and corrections $\text{R}_n$ that begin at three loops \cite{Gardi:2009qi,Becher:2009cu,Becher:2009qa}. Explicitly,
\begin{equation}
\label{eq:IRQCD}
    \Gamma_n\left( \frac{p_i}{\lambda}, \alpha_s(\lambda,\epsilon)\right)=\Gamma^{\text{dipole}}_n\left( \frac{s_{ij}}{\mu^2}, \alpha_s(\mu)\right) + \text{R}_n \, .
\end{equation}
We will consider expansions up to two loops so that there are no correction terms to worry about. The contribution of dipole terms can be written as \cite{Gardi:2009qi,Becher:2009cu,Becher:2009qa}
\begin{equation}
    \Gamma^{\text{dipole}}_n\left( \frac{s_{ij}}{\mu^2}, \alpha_s(\mu)\right) = \frac{1}{2}\hat{\gamma}_k(\alpha_s(\mu)) \sum_{i=1}^n \sum_{j=i+1}^n \log\left( \frac{-s_{ij}+i\eta}{\mu^2} \right) \mathbf{T}_i \cdot \mathbf{T}_j - \sum_{j=1}^n\gamma_j(\alpha_s(\mu)) \, , \label{eq:Gammadipole}
\end{equation}
where the color operators $\mathbf{T}_i$ carry an adjoint index. $\eta$ selects a contour prescription- we use this notation to avoid confusion with the dimensional regulator $\epsilon$. $\mathbf{T}_i$
satisfy
\begin{equation}
    [\mathbf{T}_i^a, \mathbf{T}_i^b] = i f^{ab}_{~~c} \mathbf{T}^c_i \, , \quad \mathbf{T}_i \cdot \mathbf{T}_i = \mathbf{T}_i^a \mathbf{T}_i^b \delta_{ab} = C_i^{(2)} \,  ,
\end{equation}
where $C_i^{(2)}$ is the quadratic Casimir eigenvalue of the $SU(N)$ algebra in the representation of particle $i$. When acting on the finite part $\mathcal{H}_n$, the color operators satisfy the color conservation constraint,
\begin{equation}
    \sum_{i=1}^n \mathbf{T}_i = 0 \, .
\end{equation}
In (\ref{eq:Gammadipole}), the first term depends on the cusp anomalous dimension which is representation independent up to three loops, while the second term of the same equation contains  a sum over the collinear anomalous dimensions $\gamma_i(\alpha_s(\mu))$ associated with each particle.

\cite{Gonzalez:2021dxw,Magnea:2021fvy}  showed that using the celestial coordinates and color conservation, the dipole formula factorizes nicely into a color singlet part and a color correlated part. As with the QED and gravity examples from earlier, this is due to the structure of the $s_{ij}$ dependence of the color-dipole~\eqref{eq:Gammadipole}. $s_{ij}$ itself factorizes into an $\omega$ dependent term and a $z$ dependent term, from which the log then separates. We have
\begin{equation}
    \Gamma^{\text{dipole}}_n = \hat{\Gamma}_n^{\text{corr}}(z_{ij}, \alpha_s(\lambda,\epsilon)) + \hat{\Gamma}^{\text{single}}_n(\omega_i, \alpha_s(\lambda,\epsilon)) \, ,
\end{equation}
where
\begin{subequations}
\begin{align}
    \hat{\Gamma}_n^{\text{corr}}&=-\frac{1}{4}\hat{\gamma}_k \sum_{i=1}^n\sum_{j=i+1}^n \log(|z_{ij}|^2) \mathbf{T}_i \cdot \mathbf{T}_j \, ,
    \\
    \hat{\Gamma}_n^{\text{single}} &= -\frac{1}{4} \hat{\gamma}_k \sum_{i=1}^n \log\left(\frac{\omega_i^2}{\mu^2}\right) C_i^{(2)} + D(\alpha_s(\lambda,\epsilon)) \, .
\end{align}
\end{subequations}
$D(\alpha_s(\lambda,\epsilon))$ denotes the terms that do not depend on the celestial coordinates $\omega_i$ and $z_i$.

Using the equations above in (\ref{eq:Zndef}), one finds
\begin{equation}
    \mathcal{Z}_n = \exp\left[-K \sum_{i=1}^n \sum_{j=i+1}^n \log(|z_{ij}|^2) \mathbf{T}_i\cdot \mathbf{T}_j \right] \prod_{i=1}^n \left( \frac{\omega_i}{\mu}\right)^{\frac{1}{2}K C_i^{(2)}} N_n \, , \label{eq:Zn1}
\end{equation}
where
\begin{equation}
    K=-\frac{1}{2}\int_0^{\mu^2} \frac{d\lambda^2}{\lambda^2} \hat{\gamma}_k(\alpha_s(\lambda,\epsilon)) \, .
\end{equation}
$N_n$ is a constant that accounts for the contribution of $D(\alpha_s(\lambda,\epsilon))$. Note that the first exponential factor is independent of the representation. Barring normalization $N_n$, the equation above resemblances both to the QED and gravity examples discussed earlier. Specifically,~(\ref{eq:Zn1}) implies that the Carrollian amplitudes in Yang-Mills theory also have a factorized form,
\begin{equation}
    C(\Delta_i, u_i,z_i,\bar{z}_i) = \exp\left[-K \sum_{i=1}^n \sum_{j=i+1}^n \log(|z_{ij}|^2) \mathbf{T}_i\cdot \mathbf{T}_j \right] \widetilde{C}_{\mathrm{mod}}\left(1+\frac{1}{2}K C_i^{(2)}, u_i, z_i, \bar{z}_i\right) \, , \label{eq:YMfactorized}
\end{equation}
where $\widetilde{C}_{\mathrm{mod}}$ is the modified Mellin transform of $\mathcal{H}_n$
\begin{equation}
    \widetilde{C}_{\mathrm{mod}} =  \int \prod_j \frac{d\omega_j}{2\pi} e^{i\epsilon_j \omega_j u_j} \left( \frac{\omega_i}{\mu}\right)^{\frac{1}{2}K C_i^{(2)}} N_n \mathcal{H}_n\,,
\end{equation}
with normalization $\mu^{-\frac{1}{2}K C_i^{(2)}} N_n$ included in the definition. For $\widetilde{C}_{\mathrm{mod}}$, the conformal dimension of each Carrollian operator is shifted to $1+\frac{1}{2}K C_i^{(2)}$. This shift in dimensions is analogous to the one we had encountered with QED- these shifts are formally divergent and can once again be rendered finite by working with the modified Mellin prescription and scaling the $\Delta$s appropriately with the cutoff. The above equation also mimics our story with gravity amplitudes since the soft factor is not a c-number (as was the case with QED), but an operator-albeit one that acts on the color indices rather than on the kinematic variables. As with our earlier examples, we hence see that we can render the Carrollian amplitudes IR-safe by stripping off the soft part above.

We have only considered the dipole contributions to the soft anomalous dimension. It is natural to wonder whether the structures we have uncovered survive higher loop corrections (as quantified by the R$_{n}$s in~\eqref{eq:IRQCD}). In fact, the R$_{n}$ are known to be scale invariant~\cite{Agarwal:2021ais} and consequently depend only on ratios of Mandelstams. At four points (and on-shell), this implies that these corrections depend only on the `celestial' coordinates $z_i,\overline{z}_{i}$ and not on the lightcone energies $\omega_i$. For this reason, the expectation is that the $\omega$ independent object $\hat{\Gamma}_n^{\text{corr}}$ receives modifications from three loops onwards, while the $\omega$ dependent $\hat{\Gamma}_n^{\text{single}}$ remains unaffected. As a result, we can still run through our earlier procedure of renormalizing scaling dimensions and stripping off a (modified) `soft factor' to obtain an IR finite Carrollian amplitude.


\section{Discussion} \label{sec:5}


We would like to mention a few interesting future directions related to our work. We have examined various examples of Carrollian loop amplitudes in section \ref{sec:3}. We found that at loop level, Carrollian amplitudes generically behave logarithmically. We started a study of the discontinuities of the Carrollian loop amplitudes with a specific example. It would be interesting to study discontinuities of Carrollian amplitudes more systematically. We expect that one can apply the Landau analysis \cite{Fevola:2023kaw,Fevola:2023fzn,Caron-Huot:2024brh} of loop amplitudes to Carrollian amplitudes. 

It would be interesting to study how the Carrollian OPEs are corrected by the loop corrections. Similar analysis of celestial amplitudes has been performed in \cite{Bhardwaj:2022anh,Bittleston:2022jeq,He:2023lvk,Krishna:2023ukw,Bhardwaj:2024wld,Magnea:2025zut,Banerjee:2026keq}. As mentioned in \cite{Nguyen:2025sqk}, Carrollian OPEs have a rich structure due to the presence of different branches. It would be important to check this at the loop level.

Carrollian holography is closely related to the flat limit of AdS/CFT. It has been shown that at the level of the holographic correlators, the flat limit in the bulk is realized by the Carrollian limit at the boundary \cite{Alday:2024yyj,Bagchi:2023fbj,Lipstein:2025jfj}. See, e.g. \cite{Alday:2024yyj,Surubaru:2025fmg,Kulkarni:2025qcx,Adamo:2025bfr} for examples at tree level. It would be important to show how to reproduce the Carrollian loop amplitudes in our work by taking the flat space limit of AdS Witten diagram in position space. We leave it for future work. It would also be interesting to make a connection with the other approaches for the S matrix in the spirit of flat holography recently developed in \cite{Kim:2023qbl,Kraus:2024gso,Kraus:2025wgi,Isen:2026xoc,Berenstein:2025tts,Berenstein:2025qhb,deGioia:2024yne,Navarro:2025xln}. 

As we mentioned earlier, understanding the analytic structure of Carrollian amplitudes has been an active field. One main obstruction of applying standard bootstrap methods to Carrollian amplitude is the appearance of the contact term at low point amplitudes. One possible resolution is to consider amplitudes in backgrounds with broken symmetries. See, e.g. \cite{Ruzziconi:2024zkr}. It would be interesting to study more examples with analytic properties further and compare them to the celestial analogs shown in \cite{Fan:2022vbz,Casali:2022fro,deGioia:2022fcn,Gonzo:2022tjm,Stieberger:2022zyk, Melton:2022fsf,Banerjee:2023rni,Stieberger:2023fju,Ball:2023ukj}.




\section*{Acknowledgements}
We would like to thank Zehao Zhu and Enrico Zunino for useful conversations. We would also like to thank Tim Adamo for helpful comments on the draft. BZ is supported by the Fundamental Research Funds for the Central Universities (010-63263123). VN is supported by a University of Edinburgh School of Mathematics Studentship.

\bibliographystyle{JHEP}
\bibliography{cope.bib}

\end{document}